\documentclass[final]{aipproc}

\layoutstyle{6x9}

\newcommand{\x}{$\langle x \rangle^{u-d}$}
\newcommand{\Fig}[1]{Fig.~#1}
\newcommand{\Eq}[1]{Eq.~(#1)}

\begin{document}

\title{Status of Average-$x$ from Lattice QCD}


\author{Dru B.\ Renner}{
address={Jefferson Lab, 12000 Jefferson Avenue, Newport News, VA 23606, USA}}

\begin{abstract}
As algorithms and computing power have advanced, lattice QCD has
become a precision technique for many QCD observables.  However, the
calculation of nucleon matrix elements remains an open challenge.  I
summarize the status of the lattice effort by examining one observable
that has come to represent this challenge, average-$x$:\ the fraction of
the nucleon's momentum carried by its quark constituents.  Recent
results confirm a long standing tendency to overshoot the
experimentally measured value.  Understanding this puzzle is essential
to not only the lattice calculation of nucleon properties but also the
broader effort to determine hadron structure from QCD.
\end{abstract}

\maketitle

\section{Introduction}

Understanding hadron structure from first principles is a fundamental
goal of lattice QCD.  The nucleon plays a special role as a benchmark
for lattice calculations due to the extensive experimental effort to
measure its properties.  Successfully reproducing the measured values
of basic observables, like the axial charge measured in neutron beta
decay or the charge radius measured in elastic electron scattering,
would provide a strong validation of the lattice technique.  This
would not only give confidence in the calculations of the many other
properties of the nucleon but it would also bolster the lattice effort
to calculate hadron structure more generally.  Furthermore, there is a
burgeoning program to calculate nuclear properties using lattice QCD.
It is clearly essential to have a well-controlled calculation of the
single nucleon state in order to trust future computations of
multi-nucleon systems.  Thus in many ways the nucleon is the keystone
for a much broader lattice QCD program to understand the properties of
hadrons as predicted from the underlying theory of QCD.

I have chosen to illustrate the status of the lattice effort to
understand nucleon structure by focusing on average-$x$.  This quantity
has persistently come out too high from lattice calculations.  A
variety of explanations have been offered over the years, and I'll
mention a few, but recent calculations have dramatically confirmed
this trend.  The apparent disagreement with experiment is a real
puzzle and its resolution will likely require a concentrated effort to
carefully examine all sources of error in the lattice calculation.
The advantage of using lattice QCD, as opposed to any other technique,
to calculate average-$x$ is that the list of possible errors is finite
and each source of error can be systematically removed.  This is both
a challenge and an opportunity for lattice QCD.

\section{Average-$x$}

By average-$x$, I mean specifically the difference of the up and down
contributions.  Written as a moment of the nucleon parton
distributions, average-$x$ is
\begin{equation}
\label{moment}
\langle x \rangle^{u-d}_\mu = \int_{0}^{1}\!\! dx\,\, x\, (u(x,\mu) - d(x,\mu)) + \int_{0}^{1}\!\! dx\,\, x\, (\overline{u}(x,\mu) - \overline{d}(x,\mu))\,.
\end{equation}
The unpolarized quark and anti-quark distribution functions $q(x,\mu)$
for $q=u$, $d$, $\overline{u}$ and $\overline{d}$ are extracted from
the results of many experiments, particularly deeply inelastic
electron-nucleon scattering.  It is important to remember that the
various $q(x,\mu)$ are indirectly determined by performing global fits
to the measured cross-sections and that the values of $x$ are limited
by the kinematics of each experiment.  For this reason, it would be
preferable to calculate $q(x,\mu)$ directly as a function of $x$, but
this is not possible with the lattice QCD methods that we currently
have.  This is because lattice computations are performed in Euclidean
space whereas the distributions $q(x,\mu)$ are related to the square of the light-cone
wave function, which are not easily accessible outside of Minkowski
space-time.  However, moments of the quark distributions can be
related to matrix elements of local operators, and these are
calculable in Euclidean space.  Thus lattice computations determine
$\langle x \rangle^{u-d}_\mu$ from
\begin{equation}
\label{me}
\langle p, s | \left. \left( \overline{u} \gamma_{\{ \mu } iD_{\nu \}}u - \overline{d} \gamma_{\{ \mu } iD_{\nu \}}d \right)\right|_\mu | p, s \rangle = 2 \langle x \rangle^{u-d}_\mu p_{\{\mu} p_{\nu\}}\,.
\end{equation}
As we begin to contemplate what calculations are required for a
definitive determination of average-$x$, it is important to keep in mind
that there is a significant difference between what is computed and
how that is measured.  Currently, the burden is on the lattice
community to nail down the calculations of the nucleon matrix
elements, but it is not inconceivable that there may ultimately be
some subtlety in the comparison of \Eq{\ref{moment}} and
\Eq{\ref{me}}.

For simplicity, in the following the renormalization scale $\mu$ will
be dropped and $\langle x\rangle^{u-d}$ will be understood as
evaluated in the $\overline{MS}$-scheme with $\mu=2~\mathrm{GeV}$.

\section{Persistent puzzle}

\begin{figure}
\includegraphics[width=0.85\textwidth,angle=0]{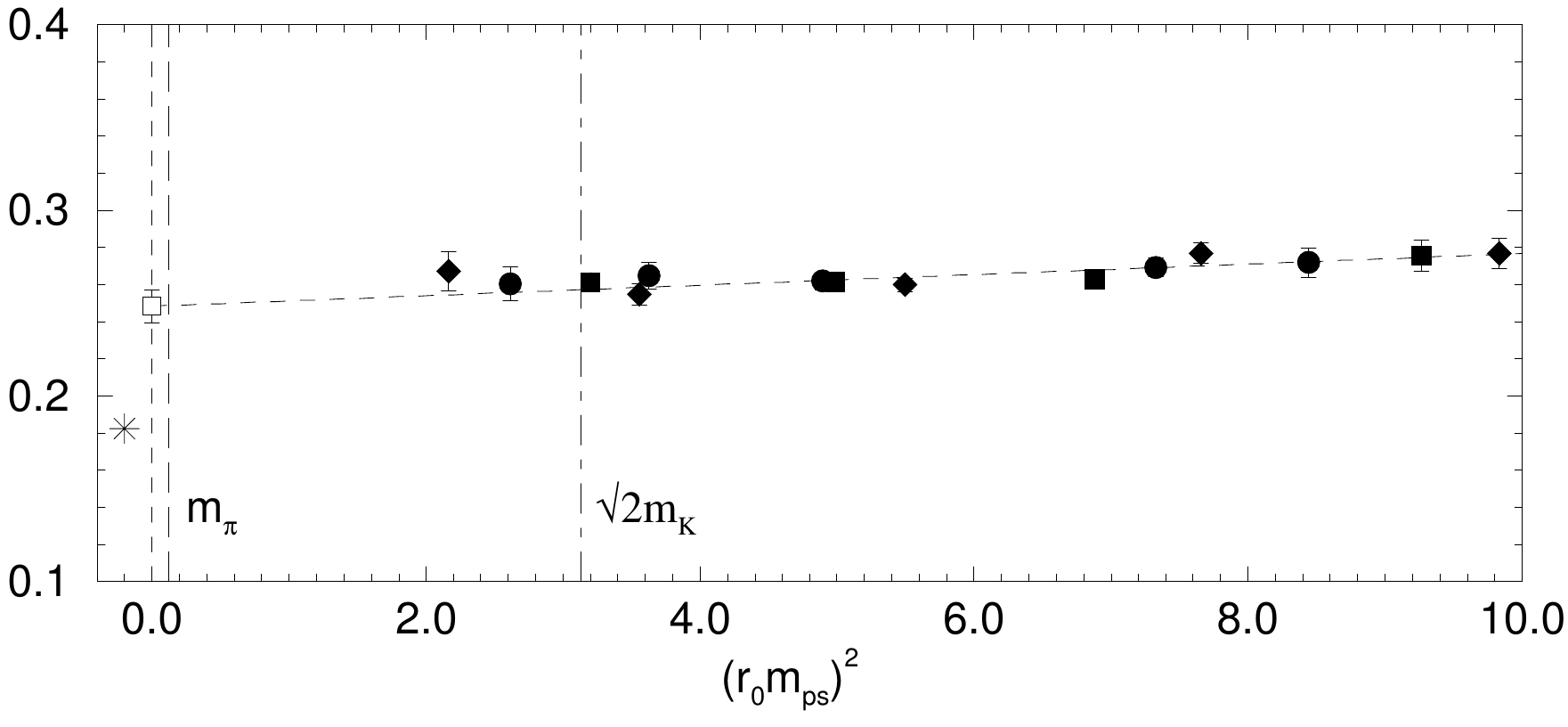}
\label{qcdsfx}
\caption{Example quenched results for {\x}.  The quenched results for
  {\x} from~\cite{Gockeler:2002ek} are plotted against the square of
  the pion mass $m_{PS}$ in units of the inverse Sommer scale
  $r_0^{-1}$.  The combination $(r_0 m_{PS})^2$ is proportional to the
  quark mass in some units as the chiral limit is approached.
  The quenched results for {\x} have a quite mild quark mass,
  equivalently pion mass, dependence over a large range of pion
  masses.  Additionally, the linear extrapolation of this pion mass
  dependence results in a substantial overestimate of the
  phenomenologically determined value for {\x}.  The quenched
  approximation was a potential source of this discrepancy that has
  since been eliminated.  This plot was taken
  from~\cite{Gockeler:2002ek}.}
\label{quenched}
\end{figure}

The puzzle with {\x} began with the earliest quenched lattice
calculations.  As an example, in \Fig{\ref{quenched}}
I show a quenched calculation of average-$x$
from~\cite{Gockeler:2002ek}.  As seen there, {\x} has a mild, nearly
flat, pion mass dependence.  Naively, this is not entirely unexpected.
Dimensionless quantities like {\x} tend to have a weaker dependence on
the quark mass than dimensionful quantities like the nucleon mass.  In
fact, this sort of behavior would normally be welcomed, except in this
case the lattice calculation of {\x} clearly overshoots the
phenomenologically determined value.

\begin{figure}
\includegraphics[width=0.75\textwidth,angle=0]{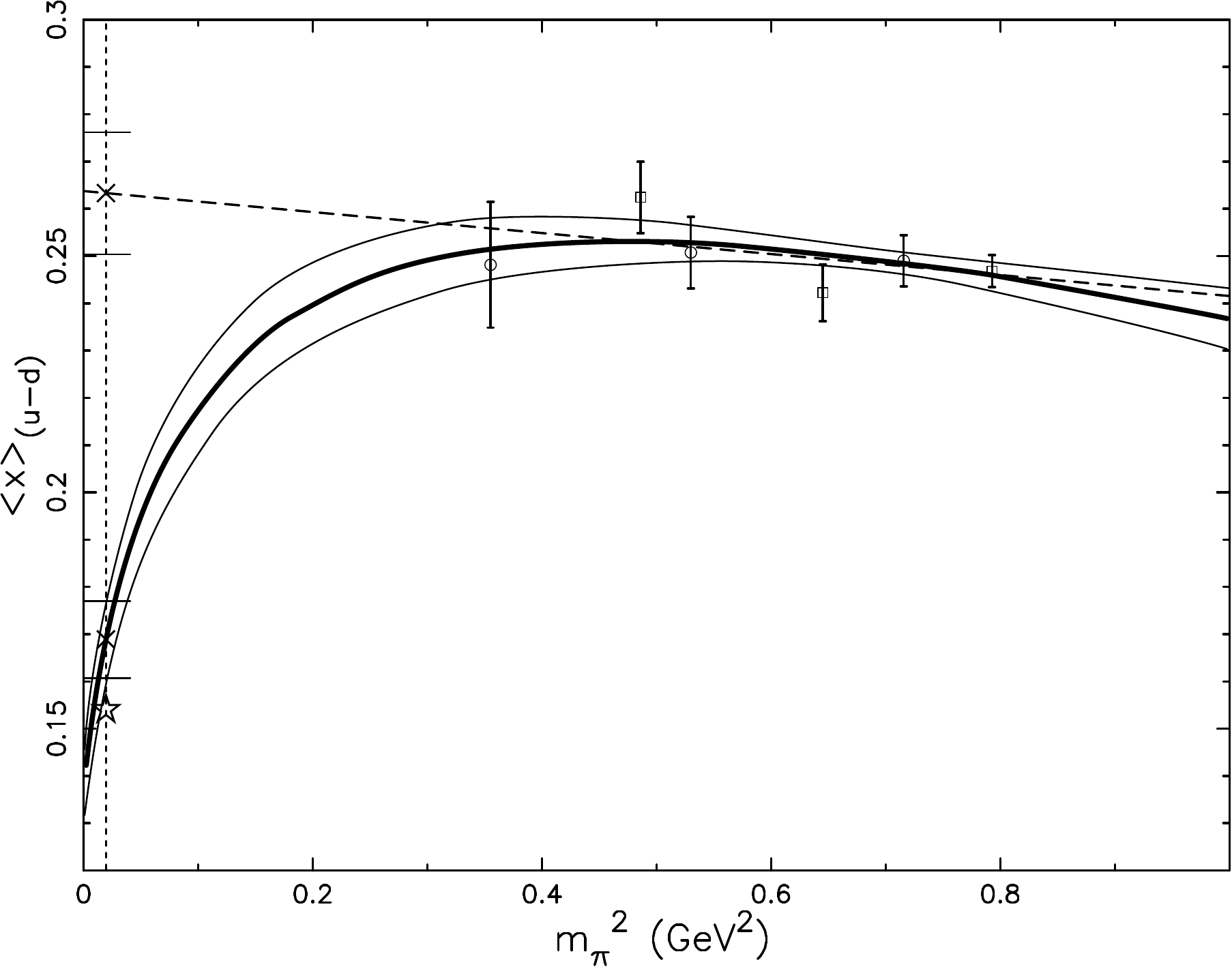}
\caption{First full QCD results for {\x}.  The results for {\x}
  from~\cite{Dolgov:2002zm} are shown.  They are plotted against the
  square of the pion mass.  Similar to the quenched results, average-$x$
  from full QCD calculations, like the one shown here, have a mild
  pion mass dependence and overshoot the phenomenological value for
  {\x}.  Chiral perturbation theory predicts the leading pion mass
  dependence of {\x}~\cite{Arndt:2001ye,Chen:2001eg}.  When combined
  with a physically motivated regulator~\cite{Detmold:2001jb}, the
  resulting functional form was capable of smoothly matching the
  lattice computation to the expected pion mass dependence in the
  chiral limit.  Thus it was hypothesized that chiral dynamics might
  help explain the seemingly strong quark mass dependence that would
  be required to make the lattice results agree with the physical
  value of {\x}.  This explanation is being challenged by current
  calculations.  This plot was taken from~\cite{Dolgov:2002zm}.}
\label{chiral}
\end{figure}

At the time it was natural to dismiss this problem as simply being an
artifact of the quenched approximation that drops all contributions
from the so-called sea-quark loops.  However, the results from the
earliest full QCD calculation~\cite{Dolgov:2002zm}, which included two
dynamical quark flavors, were found to agree with the quenched
calculations.  The two-flavor results from~\cite{Dolgov:2002zm} are
shown in \Fig{\ref{chiral}},
where again the lattice calculations came out too high.

Staying with \Fig{\ref{chiral}}, another explanation for the puzzling
behavior of {\x} was put forth.  The pion mass dependence of average-$x$
was calculated in chiral perturbation theory~\cite{Arndt:2001ye,
  Chen:2001eg} and combined with a phenomenological
regulator~\cite{Detmold:2001jb} that was capable of accommodating both
the lattice calculation and the physical value of {\x}.  It was
understood that the pion masses were too heavy to apply chiral
perturbation but~\cite{Detmold:2001jb} offered a plausibility argument
that chiral dynamics may lead to a strong quark mass dependence for
yet lighter quark masses while producing a mild quark mass dependence
for the range of quark masses used in contemporary lattice
computations.  This was put on a slightly stronger footing with
calculations to higher-order in the chiral
expansion~\cite{Dorati:2007bk}.  It was shown that appropriate choices
of the undetermined counterterms in the resulting functional form
could lead to a flat pion mass dependence for heavy pion
masses~\cite{Dorati:2007bk, Hagler:2007xi}.

This line of reasoning has dominated the lattice effort on nucleon
structure for much of the last decade.  It was understood that
physically motivated regulators would introduce model dependence to
the extrapolation of the lattice calculations.  It also seemed that
higher-order calculations would require the determination of too many
extra counterterms and low energy constants to be practically useful.
Thus the hope was to push to light enough pion masses to directly
observe the missing \emph{chiral logarithms}.  These are the
contributions to {\x} of the form $m_\pi^2 \ln m_\pi^2$ that are, more
or less, uniquely predicted by chiral perturbation theory.

Fully dynamical lattice calculations of {\x} have continued to lighter
quark masses in search of these missing logarithms.  The initial
two-flavor calculations~\cite{Dolgov:2002zm, Lin:2008uz, Baron:2008zz,
  Pleiter:2011gw} have been extend to include the strange
quark~\cite{Ohta:2010sr, Hagler:2007xi, Winter:2011ze} and extended
further to even include the charm quark~\cite{Dinter:2011jt}.  The
basic observation from the earliest quenched calculations remains
correct:\ average-$x$ appears to have a mild pion mass dependence, the
extrapolation of which is higher than the physical value.  The
lightest pion mass used in the calculations referenced so far was
approximately $250~\mathrm{MeV}$.  Understanding that finite-size
effects and lattice artifacts are seldom checked at the lightest pion
mass used in a calculation, we could argue that the lightest reliable
pion mass was closer to $300~\mathrm{MeV}$.  This left some room for
the rapid pion mass dependence that would be required to reconcile the
lattice computation with the experimental measurement of {\x}, but
recent calculations have begun to challenge this scenario.

\section{Recent results}

\begin{figure}
\includegraphics[width=0.8\textwidth,angle=0]{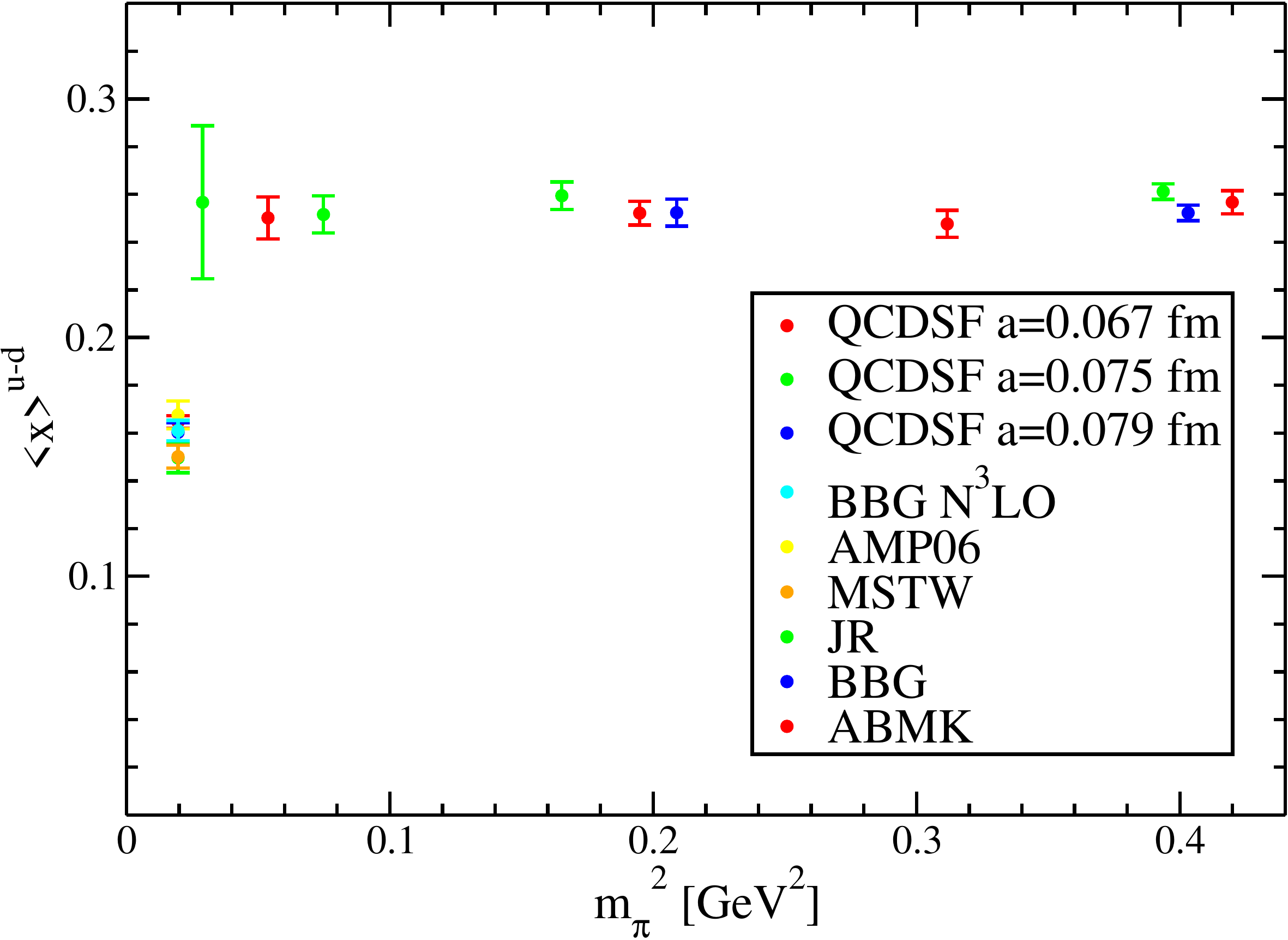}
\caption{Recent results from the QCDSF collaboration for {\x}.
  Preliminary results from the QCDSF
  collaboration~\cite{Pleiter:2011gw} for {\x} are plotted versus the
  square of the pion mass.  These results rather dramatically continue
  the long trend of lattice calculations of {\x} with quite mild pion
  mass dependence that extrapolates to values noticeably higher than
  the experimental measurement.  These results are challenging the
  prevailing view that chiral dynamics would cause sufficient
  curvature in $m_\pi^2$ to reconcile the lattice calculations at
  heavy pion masses with the value of {\x} at the physical point.
  Also shown are a set of the most recent results for {\x} from global
  analyses.  These were collected in~\cite{Renner:2010ks} using
  results from~\cite{Alekhin:2009ni, Blumlein:2006be, Blumlein:2004ip,
    JimenezDelgado:2008hf, Martin:2009bu, Alekhin:2006zm}.  Note that all these results are N${}^2$LO except
    for the one explicitly marked as N${}^3$LO.  The
  results in this plot were communicated privately by the QCDSF
  collaboration.}
\label{qcdsf}
\end{figure}

Many of the most recent results were summarized
in~\cite{Alexandrou:2010cm}, but rather than showing all the
calculations of average-$x$, I focus on the results from the QCDSF
collaboration~\cite{Pleiter:2011gw}.  Their calculation of {\x}
illustrates the most recent trend in lattice calculations:\ several
collaborations are now calculating at or near the physical pion
mass~\cite{Ohta:2011nv, Durr:2010aw, Aoki:2009ix}.  There are a
variety of compromises that are made to accomplish this, but it is
still a very important advance.  The calculation of {\x}
from~\cite{Pleiter:2011gw}, with pion masses approaching the physical
pion mass, is shown in \Fig{\ref{qcdsf}}.
It is very plain to see that these latest results for average-$x$
confirm the nearly flat pion mass dependence of {\x} down to
essentially the physical point.  This calculation achieves a long
sought milestone, but the conclusion is far from clear.

If the results from~\cite{Pleiter:2011gw} are taken at face value,
then it is hard to escape the obvious conclusion that one would draw
from \Fig{\ref{qcdsf}}.  Either there are unaccounted for sources of
error in the lattice computation (or the global fits) or there is a
sizable discrepancy between the lattice determination of {\x} and the
experimental measurement of it.  This later option seems unlikely, so
the current view among those doing the lattice calculations of {\x} is
that one or more of the systematic errors that must be checked for in
lattice calculations is currently underestimated.

Regarding the possibility of underestimated errors in the value of
{\x} extracted from the experimental measurements, I have shown the
results from six recent analyses of average-$x$ in \Fig{\ref{qcdsf}}.
There is some spread beyond that expected by the quoted errors on
{\x}, but it is certainly not large enough to account for the
difference between the lattice results and the global fits.  Thus it
seems unlikely that there is a significant problem in the
phenomenological results for {\x}, but it is useful to keep in mind
that an extrapolation in $x$ is required to evaluate the integral in
\Eq{\ref{moment}}.  Additionally, lattice calculations
often fail to specify the order to which the matching to $\overline{MS}$ is done,
but the difference between the experimental N${}^2$LO and N${}^3$LO results
in \Fig{\ref{qcdsf}}
suggests that this effect is small.  

\section{Systematic errors}

The resolution of the puzzle in \Fig{\ref{qcdsf}} will likely hinge on
a careful examination of all the systematic errors present in the
lattice calculation of {\x}.  Most of these sources of uncertainty
have been checked, to some extent, in previous calculations, so it was
believed that the dominant source of error in {\x} was due to the
poorly constrained extrapolation in the pion mass.  However, the
results in \Fig{\ref{qcdsf}} now suggest that this might not be the
case.  Certainly, the chiral extrapolation is no longer the single
stand out systematic error.  This raises the possibility that other
errors were underestimated or that the discrepancy in
\Fig{\ref{qcdsf}} could be a combination of several smaller
uncertainties.

The advantage of using a renormalizable description of the fundamental
theory is that we know with confidence that the list of potential
errors is very limited.  First, we have to reliably calculate the
basic nucleon matrix element in \Eq{\ref{me}}.  This involves the
underlying algorithms used to stochastically evaluate the functional
integrals that define the matrix element.  Since these methods are
used in many successful lattice computations, it seems unlikely that
there is a special algorithmic problem for the nucleon.  Calculating
the matrix element in \Eq{\ref{me}} also requires isolating the ground
state, corresponding to the nucleon, using Euclidean space methods.
Several ongoing investigations~\cite{privQCDSF,privETMC} suggest that
this may be responsible for some, but not all, of the discrepancy in
{\x}.  This can be called the \emph{plateau problem} because most
calculations of nucleon matrix elements rely on finding a plateau as a
function of Euclidean time in appropriately chosen correlation
functions.  This issue has been examined off and on, most recently
in~\cite{Capitani:2010sg}, and a variety of new methods have been
developed to address it~\cite{Blossier:2009kd, Foley:2010vv}.

Once we have calculated the so-to-speak bare matrix element, the
operator renormalization must be accounted for.  This is now regularly
calculated using nonperturbative methods, thus eliminating one
potential source of error.  However, the method for renormalizing
composite operators nonperturbatively has its own set of potential
systematic errors.  Most of these are quite technical in nature and go
well beyond the level of presentation in these proceedings, but the
overarching concern regards the separation of scales that is necessary
to nonperturbatively match the lattice operator to the continuum
$\overline{MS}$ operator needed in \Eq{\ref{me}}.  Because
$\overline{MS}$ is a perturbative scheme, the matching must ultimately
involve some form of perturbation theory.  To reduce the error from
this, the matching is done, ideally, at large renormalization scales
$\mu$, but this runs afoul of the constraint $\mu \ll 1/a$ that must
be maintained to control lattice cut-off effects.

There is some indirect evidence for a problem in the renormalization.
In~\cite{Renner:2010ks} it was pointed out that there does appear to
be some variation in the relative normalization of the results for
{\x} from different actions.  Additionally, it was be found that
ratios of observables that cancel the renormalization lead to results
in agreement with experimental measurements~\cite{Pleiter:2011gw,
  Ohta:2011nv}.  Additionally, the renormalization is an
interesting potential culprit because it would lead to a simple
rescaling of {\x}.  This is because $\overline{MS}$ is a mass
independent scheme, so the renormalization of the operator depends
only on the lattice spacing and not the quark masses.  Thus a mistake
in the renormalization would correspond to a multiplicative rescaling
of \Fig{\ref{qcdsf}}, for example.  It seems unlikely that such an
effect would account for the entire discrepancy, but it may be one
piece of the puzzle.  As a separate cross check, there are attempts to
eliminate the renormalization issue entirely and calculate moments of
structure functions directly~\cite{Bietenholz:2010kn}.

Having reliably calculated the properly renormalized matrix element,
then the only remaining systematic errors are the extrapolation of the
heavier-than-physical pion mass to the physical pion mass and the
continuum and infinite-volume limits.  Collectively, the results from
all the calculations of {\x} suggest that these errors are small
within the range of pion mass, lattice spacing and volumes that have
been used.  The concern, though, is that the asymptotic values of each
of the three limits may not have been seen in the currently used
ranges.  The size of lattice artifacts can be checked by establishing
not just weak lattice spacing dependence, as is customary, but by
demonstrating the expected scaling as the continuum limit is
approached.  For the $L$ dependence, one could explicitly check the
generically expected exponential suppression at large $L$, rather than
the current standard of simply demonstrating an apparent convergence
to within the errors.  It is hard to point to any compelling
indications of finite-size or cut-off effects in the current
calculations of {\x}, but that may be so because these issues have
never been pursued with the high precision likely required to detect
such effects.

The pion mass dependence is harder to check because the only
expectations come from chiral perturbation theory and that may simply
not be applicable for the physical pion mass or heavier.  But as
\Fig{\ref{qcdsf}} demonstrates, current calculations are quickly
reducing the extent of the required extrapolation in the pion mass and
not-so-far-off calculations will be able to bridge the physical pion
mass, thus converting an extrapolation into an interpolation.

We also must keep in mind that each of these possible systematics can
interfere with each other.  For example, failing to reliably determine
the matrix element may produce results that erroneously have a flat
pion mass dependence or failing to properly renormalize the needed
operator may obscure the cut-off dependence.  And of course, all the
calculations must ultimately be done with sufficient statistical
precision to be capable of clearly checking the, relatively short,
list of systematic errors.

\section{Conclusions}

Lattice calculations are proceeding steadily down to the physical pion
mass.  This has facilitated the precision calculation of many QCD
observables, however, it has also produced some puzzles.  In
particular, recent calculations of average-$x$ have continued a long
established trend of overshooting the value for {\x} determined by
global analyses.  For quite some time, this had been assumed to be
caused by a suppression of chiral dynamics due to the use of
heavier-than-physical pion masses, but this explanation is much less
compelling in the light of recent results.  Clarification of this
situation will likely require a careful study of all the possible
uncertainties in the computation of {\x}.  The use of a well-defined
nonperturbative regulator, namely lattice QCD, ensures that the
errors in the calculation of {\x} are identifiable and all
systematically improvable.  Controlling the chiral limit is still, of
course, one source of error in the lattice calculations, but it is now
just one among several potentially comparable errors.  This marks a
milestone in the lattice effort to determine hadron structure directly
from QCD.

\begin{theacknowledgments}
This work was supported by SciDAC and by Jefferson Science Associates,
LLC under U.S. DOE Contract No. DE-AC05-06OR23177. The U.S. Government
retains a non-exclusive, paid-up, irrevocable, world-wide license to
publish or reproduce this manuscript for U.S. Government purposes.
\end{theacknowledgments}

\bibliographystyle{aipproc} 

\bibliography{lat-hix}

\end{document}